\documentstyle[12pt,epsfig]{article}
\begin{document}
{\Large\bf Leading/nonleading charm production asymmetry in $\Sigma^-p$
interactions.}
\\[2mm]

{\it O.I.Piskounova}\\[2mm]

{\small P.N.Lebedev Physical Institute of Russian Academy of Science, 
Moscow}
\vspace{1cm}
\begin{abstract}

The asymmetries between the spectra of leading and nonleading charmed
mesons
measured in $\Sigma^-A$ interactions at $p_L$= 340 GeV/c in the
WA89 experiment are described in the framework of Quark-Gluon String Model.
There are two versions of the model under consideration: one of them
includes
the sea charm quark-antiquark pairs and the other one does not. It's shown 
that the asymmetries between $D^-$ and $D^+$-meson spectra and between 
$D_s^-$ and $D^+_s$- meson spectra can be fitted by QGSM curves obtained 
with the same parameters as charm asymmetry
in $\pi^-A$ experiments described in previous studies. The QGSM results are
compared with the calculations in the next-to-leading approximation of 
perturbative QCD approach carried out by  the other authors.

\end{abstract}

\section{Introduction}

The asymmetries between the spectra of $D^-$ and $D^+$, $D_s^-$
and $D^+_s$  mesons were measured recently in $\Sigma^-A$ interactions 
at $p_L$= 340 GeV/c in the  WA89 experiment \cite{wa89}. 
It seems to be interesting to compare this asymmetries 
with those ones obtained in the $\pi^-A$ experiments in order to 
understand the influence of quark composition of beam particle on 
a heavy flavored particle production and to extract some specific 
features of the strange-charmed meson spectra caused by
the presence of a strange valence quark in $\Sigma^-$ hyperon.
The difference in $x_F$ spectra($x_F=2p_{II}/\sqrt{s}$) of leading and 
nonleading particles  has been discussed recently and several theoretical 
models explained successfully the asymmetry as an effect of 
an interplay between the quark contents of the projectile
and of the produced hadron. The charmed mesons containing ordinary 
quarks of the same type as beam particle
have higher average x value. The asymmetry, defined as:

\begin{equation}
A(x)=\frac{dN^{D^-}/dx-dN^{D^+}/dx}{dN^{D^-}/dx+dN^{D^+}/dx},
\end{equation}

is a function rising with x. There are two different approaches in the
theory 
for the description of this effect. The first one is based on perturbative 
theory of QCD. It takes into account the recombination of 'intrinsic charm'

quarks with valence quarks of the projectile as an origine of the asymmetry

\cite{qcd}.
The other phenomenological models exploit the properties of fragmentation 
functions in order to insert the asymmetry .
We will not discuss here the details of recombination models but
we are going to concentrate on a nonperturbative approach known as the 
Quark Gluon String Model (QGSM) \cite{qgsm}.   This model  fits well
the leading/nonleading charm asymmetry for $\pi^-p$ experiments
\cite{pionbeam}.

\section{The Quark Distributions in QGSM.}


The inclusive production cross section of D-mesons is written as 
a sum over n-Pomeron cylinder diagrams:

\begin{equation}
f_{1}=x \frac{d \sigma^{D}}{dx}(s,x)= \int E \frac{d^{3}\sigma^{D}}
{d^{3}p}d^{2}p_{\bot}=\sum_{n=0}^{\infty}\sigma_{n}(s) \varphi_{n}^{D}(s,x)
\end{equation}

where function  $\varphi_{n}^{D}(s,x)$ is a particle distribution in the
configuration of n cut cylinders and $\sigma_{n}$ is the probability of 
this process. The parameter of the supercritical Pomeron used here 
is $\Delta_P=\alpha_P(0)-1=0,12$. The detailed formulae for $\sigma_{n}$ 
and $\varphi_{n}^{D}$ in pp-interactions can be found in \cite{prev
papers}.

The distribution functions in case of $\Sigma^-$-p collisions are:
\begin{equation}
\varphi _{n}^{D}(s,x)=a_{0}^{D}({F_{q}^{(n)}(x_{+})F_{qq}^{(n)}(x_{-})+
F_{qq}^{(n)}(x_{+})F_{q}^{(n)}(x_{-})+2(n-1)F_{q_{sea}}^{(n)})
F_{\bar{q}_{sea}}^{(n)}(x_{-})},
\end{equation}
where $a_0^D$ is the density parameter of quark-antiquark chain
fragmentation
into the given type of mesons.

The particle distribution on each side of chain can be build on the 
account of quark contents of beam particle
($x_+~=~(x+\sqrt{x^2+x_{\bot}^2})/2$,
$x_{\bot}=2m_{\bot}/sqrt{s}$) and  of target particle ($x_-~=~(x-\sqrt{x^2+
x_{\bot}^2})/2$):

\begin{eqnarray}
 F_{q}^{(n)}(x_{+})&=&\frac{1}{3}F_{s}^{(n)}(x_{+})+\frac{2}{3}F_{d}^{(n)}(x
_+),\nonumber \\
 F_{qq}^{(n)}(x_{+})&=&\frac{1}{3}F_{dd}^{(n)}(x^{+})+\frac{2}{3}F_{ds}^{(n)
}(x_{+}), \\
 F_{q}^{(n)}(x_{-})&=&\frac{1}{3}F_{d}^{(n)}(x_{-})+\frac{2}{3}F_{u}^{(n)}(x
_{-}),\nonumber \\
 F_{qq}^{(n)}(x_{-})&=&\frac{1}{3}F_{uu}^{(n)}(x_{-})+\frac{2}{3}F_{ud}^{(n)
}(x_{-}).\nonumber
\end{eqnarray}

Each $F_{i}(x_{\pm})$ is constructed as the convolution:

\begin{equation}
 F_{i}(x_{\pm})=\int_{x_{\pm}}^{1}
f_{\Sigma^-}^{i}(x_{1})\frac{x_{\pm}}{x_{1}}
 {\cal D}_{i}^{D}(\frac{x_{\pm}}{x_{1}})dx_{1},
\end{equation}

where $f^{i}(x_{1})$ is a structure function of i-th quark wich has a
fraction of energy $x_{1}$ in the interacting hadron  and ${\cal
D}_{i}^{D}(z)$
is a fragmentation function of this quark into the considered type 
of  D or $D_s$-mesons.

The structure functions of quarks in interacting proton have already been
described
in the previous papers \cite{prev papers}. In the case of hyperon they
depend
on the parameter of the Regge trajectory of $\varphi$-mesons ($s\bar{s}$)
because of s-quark contained in $\Sigma^-$:

\begin{eqnarray}
f_{\Sigma^{-}}^{d}(x_{1})&=&C_{d}^{(n)}x_{1}^{-\alpha_{R}(0)}(1-x_{1})^
{\alpha_{R}(0)-2\alpha_{N}(0)+\alpha_{R}(0)+n-1},\nonumber  \\
f_{\Sigma^{-}}^{dd}(x_{1})&=&C_{dd}^{(n)}x_{1}^{\alpha_{R}(0)-2\alpha_{N}(0)
}
(1-x_{1})^{-\alpha_{\varphi}(0)+n-1}, \\
f_{\Sigma^{-}}^{ds}(x_{1})&=&C_{ds}^{(n)}x_{1}^{\alpha_{R}(0)-2\alpha_{N}(0)
+
\alpha_{R}(0)-\alpha_{\varphi(0)}}(1-x_{1})^{-\alpha_{R}(0)+n-1},\nonumber
\\
f_{\Sigma^{-}}^{s}(x_{1})&=&C_{s}^{(n)}x_{1}^{-\alpha_{\varphi}(0)}(1-x_{1})
^
{\alpha_{R}(0)-2\alpha_{N}(0)+\alpha_{R}(0)-
\alpha_{\varphi}(0)+n-1},\nonumber \\
\end{eqnarray}

where $\alpha_{\varphi}$(0)=0. The constants $C_{i}^{(n)}$ are determined
by
normalization conditions:
$$
\int_{0}^{1}f_{i}(x_{1})dx_{1}=1.
$$
\section{The Fragmentation Functions.}

The fragmentation functions are constructed for the quark and diquark 
chains according to the rules proposed in \cite{kaidalov}. The following
favoured fragmentation function into $D_{s}^{-}$-mesons
was written for the strange valence quark:

\begin{equation}
{\cal D}_{s}^{D_{s}^{-}}(z)=\frac{1}{z}(1-z)^{-\alpha_{\psi}(0)+\lambda}
(1+a_{1}^{D_{s}}z^{2}),
\end{equation}

where  $\lambda$=2$\alpha_{D^{*}}^{\prime}$(0)$\bar{p_{\bot D^{*}}^{2}}$.
An additional factor $(1+a_{1}^{D_{s}}z^{2})$
provides the parametrization of the probability of heavy
quark production in the interval z=0 to z $\rightarrow$ 1.  The values
of the constant $a_{1}^{D_{s}}$  will be discussed later.

The function for the nonleading fragmentation of d-quark chain is:

\begin{equation}
{\cal D}_{d}^{D^{+}}(z)=\frac{1}{z}(1-z)^{-\alpha_R(0)+\lambda+
2(1-\alpha_{R}(0))+\Delta_{\psi}},
\end{equation}

where $\Delta_{\psi}=\alpha_{R}(0)-\alpha_{\psi}$(0).
The function of the nonleading fragmentation of the diquark chain is
the following:

\begin{equation}
{\cal D}_{d}^{D_{s}^{-}}(z)=\frac{1}{z}(1-z)^{-\alpha_{\varphi}(0)+\lambda+
2(1-\alpha_{R}(0))+\Delta_{\psi}}
\end{equation}

where $\alpha_{R}$(0)=0.5 and
$\Delta_{\psi}=\alpha_{R}(0)-\alpha_{\psi}$(0).

The following fragmentation function corresponds to the version of the 
diquark fragmentation into $D_{s}^{-}$-mesons:

\begin{equation}
{\cal
D}_{ds}^{D_{s}^{-}}(z)=\frac{1}{2z}(1-z)^{\alpha_{R}(0)-2\alpha_{N}(0)+
\lambda+\Delta_{\psi}}(1+a_{1}^{D_{s}}
z^{2})+\frac{1}{2z}(1-z)^{\alpha_{R}(0)-
2\alpha_{N}(0)+\lambda+\Delta_{\varphi}+\Delta_{\psi}+1}
\end{equation}

\section{The Asymmetry Suppression Causes.}

Some fractions of sea quark pairs in hyperon, $d\bar{d}$ and $s\bar{s}$,
are to 
be taken into account as far as they suppress the leading/nonleading
asymmetry. 
The structure functions of ordinary
quark pairs in the quark sea of hyperon can be written by the same way 
as the valence quark distributions:

\begin{eqnarray}
f_{\Sigma^{-}}^{d}(x_{1})&=&C_{d,\bar{d}}^{(n)}x_{1}^{-\alpha_{R}(0)}(1-x_{1
})^
{\alpha_{R}(0)-2\alpha_{N}(0)+\Delta_{\varphi}+n-1+2(1-\alpha_{R}(0))},
\end{eqnarray}

where sea quarks and antiquarks have an additional power term
$2(1-\alpha_{R}(0))$ corresponding to the quark distribution 
of two pomeron diagram which includes one sea quark pair.   

The structure function for strange sea quarks obeys the same rules:

\begin{eqnarray}
f_{\Sigma^{-}}^{d}(x_{1})&=&C_{s,\bar{s}}^{(n)}\delta_{s,\bar{s}}x_{1}^
{-\alpha_{\varphi}(0)}(1-x_{1})^{\alpha_{R}(0)-2\alpha_{N}(0)+
\delta_{\varphi}+n-1+2(1-\alpha_{R}(0))}
\end{eqnarray}

where $\Delta_{\varphi}=\alpha_R(0)-\alpha_{\varphi}(0)$ and 
$\delta_{s,\bar{s}}$=0.25 ( see \cite{kaons}).

The fragmentation function of strange sea quark( or antiquark) into 
$D_s$ mesons has
the following form for mesons of the both charges:

\begin{equation}
{\cal D}_{s,\bar{s}}^{D_{s}^{-}}(z)=\frac{1}{z}z^{1-\alpha_{\varphi}(0)}
(1-z)^{-\alpha_{\varphi}(0)+\lambda+2(1-\alpha_{R}(0))+\Delta_{\psi}}
\end{equation}

The additional fragmentation  parameter $a_f^{D_s}$ is equal to the 
fragmentation  parameter for D-mesons.  

\section{The Intrinsic Charm Distribution.}

As soon as we accounted $d\bar{d}$ and $s\bar{s}$ fraction in the quark 
sea of hyperon some fraction of charmed sea quark are to be 
considered as well. This small heavy quark admixture 
plays an important role due to its strong impact on the difference 
between leading and nonleading charmed meson spectra. 

The charmed sea quark structure function is similar to the 
distribution of strange sea quarks:

\begin{eqnarray}
f_{\Sigma^{-}}^{c,\bar{c}}(x_{1})&=&C_{c,\bar{c}}^{(n)}\delta_{c,\bar{c}}
x_{1}^{-\alpha_{\psi}(0)}(1-x_{1})^{\alpha_{R}(0)-2\alpha_{N}(0)+
\Delta_{\varphi}+n-1+2(1-\alpha_{R}(0))}
\end{eqnarray}

where $\delta_{c,\bar{c}}$ is the weight of charm admixture in 
the quark sea of hyperon. In fact it is not neseccarily to be equal to 
the charmed quark fraction in quark sea of pion \cite{pionbeam}. This is 
only one parameter we
can vary for $\Sigma^{-}$ interaction after the best fit of pion
experimental data which had been done before. The value 
of $\delta_{c,\bar{c}}$ 
can be estimated in the description of the WA89 data on $D_s$ and 
D meson asymmetries.

Fragmentation functions are the following:

\begin{equation}
{\cal D}_{c,\bar{c}}^{D_{-}}(z)=\frac{1}{z}z^{1-\alpha_{\psi}(0)}(1-z)^
{-\alpha_{R}(0)+\lambda}
\end{equation}

and for $D_s$ :

\begin{equation}
{\cal D}_{c,\bar{c}}^{D_{s}^{-}}(z)=\frac{1}{z}z^{1-\alpha_{\psi}(0)}(1-z)^
{-\alpha_{\varphi}(0)+\lambda}
\end{equation}

\section{The Final Plots and The Comparisons.}

The main parameter of QGSM scheme which is responsable for 
leading/nonleading charm asymmetry is $a_1$. It is the parametrization 
parameter of leading fragmentation function dependence on z 
$\rightarrow$ 1. The fraction of charmed sea quarks,
$\delta_{(c,\bar{c})}$, 
is the second parameter in this calculations which makes the asymmetry 
lower because of the equal amounts of $D^+$ and $D^-$ mesons produced by 
each sea charmed quark pair. Two sets of this couple of parameters were
chosen in the description of $\pi^-A$ reaction data: $a_1$=4, $\delta_{(c,
\bar{c})}$=0 and  $a_1$= 10, $\delta_{(c,\bar{c})}= 0.05\sqrt{a_0^D}$. 
We concider
here these two values of $a_1$ taking the $\delta_{(c,\bar{c})} $ 
as more or less free parameter.              

\begin{figure}[h]
\begin{center}
\vspace*{-.5cm}
\epsfig{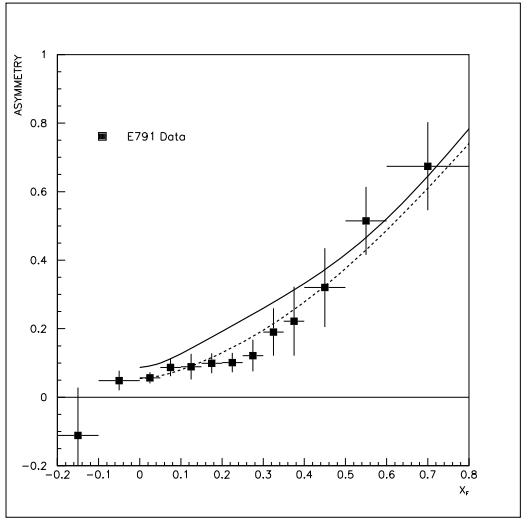}
\caption{Asymmetry between $D^-$ and $D^+$ spectra obtained in E791 
(black squares) \protect\cite{E791} and QGSM curves at two versions of 
parameters: solid line corresponds to the following set of 
parameters $a_1=10,\delta_{(c,\bar{c})}=0.05\sqrt{a_0^D}$; dashed line
is a result of QGSM fit with  $a_1=4,\delta_{(c,\bar{c})}=0$.} 
\end{center}
\end{figure}

The two curves displayed in Fig.1 represent the fits of E791 
pion beam experiment data \cite{E791} with two sets of parameters 
discussed above.
Data of the WA89 experiment are given in Fig.2 and Fig.3 
with the same parameters. It should be mentioned 
that the smaller fraction of charmed sea quarks was taken into 
account ($\delta_{(c,\bar{c})}=0.01\sqrt{a_0^D}$ for to 
describe both $D^-/D^+ and D_s^-/D^+_s$ asymmetries with 
the fragmentation parameter $a_1=10$.

 The resulting curves obtained in several theoretical models 
\cite{lichoded,arakelyan} are also shown in these Figures as well.  

\begin{figure}[h]
\begin{center}
\vspace*{-.5cm}
\epsfig{figure=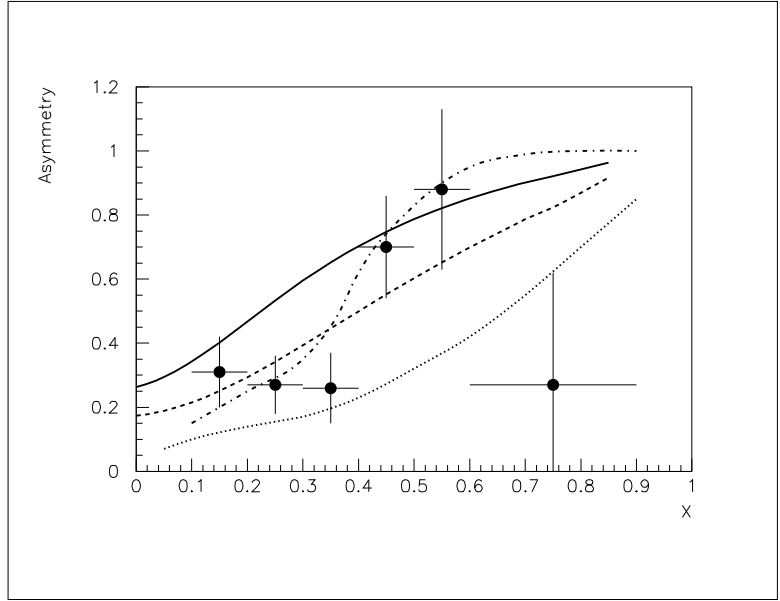,height=8cm}
\caption{$D^-/D^+$ asymmetry measured in WA89 and theoretical calculations:
solid line corresponds to the following set of QGSM parameters $a_1=10,
\delta_(c,\bar{c})=0.01\sqrt{a_0^D}$; dashed line is a result of QGSM 
fit with  $a_1=4,\delta_{(c,\bar{c})}=0$; dashed-dotted line is the result
of \protect\cite{lichoded} and dotted line corresponds to A(x) predicted 
in \protect\cite{arakelyan}.}
\end{center}
\end{figure}

\begin{figure}[h]
\begin{center}
\vspace*{-.5cm}
\epsfig{figure=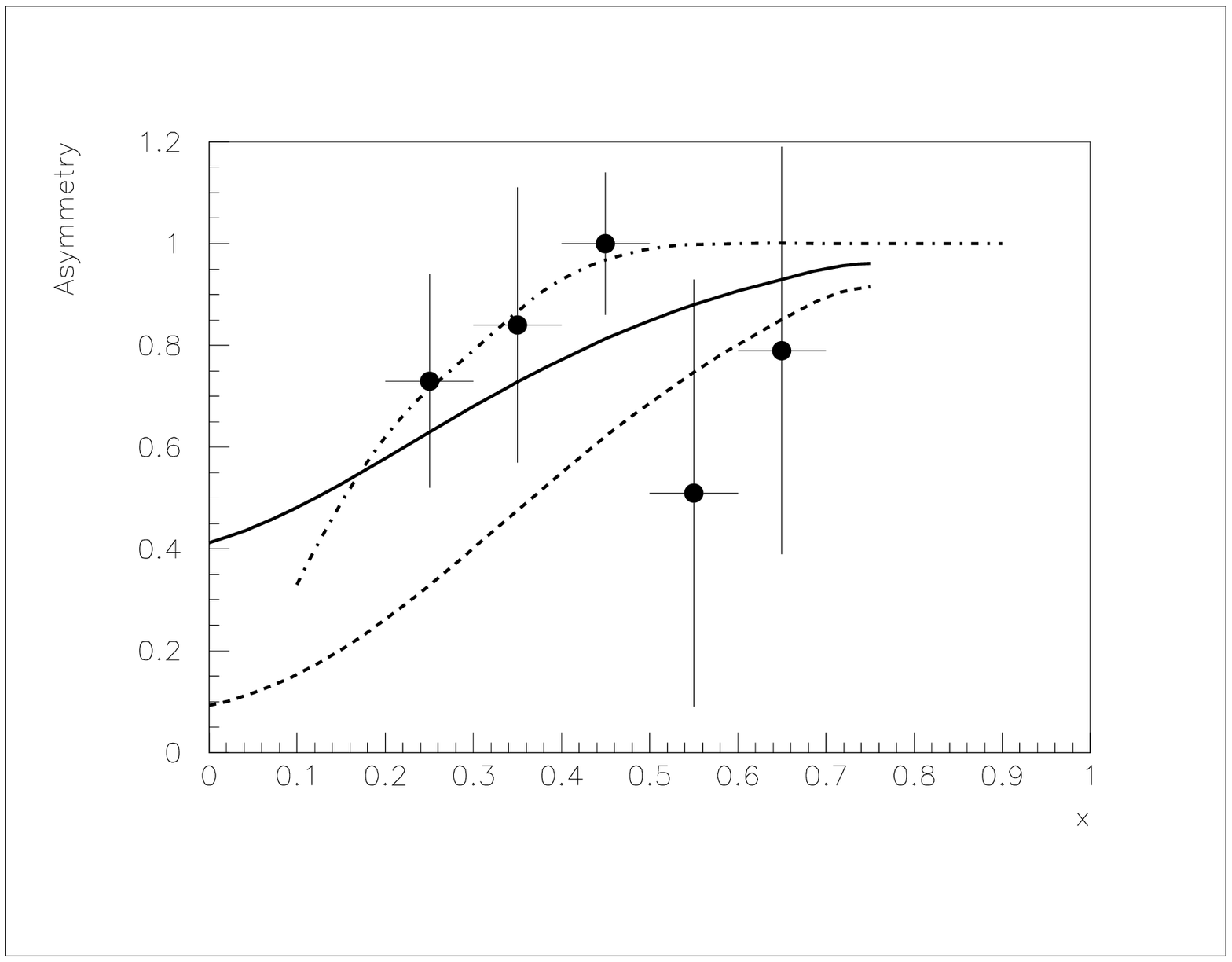,height=8cm}
\caption{$D^-_s/D^+_s$ asymmetry measured in WA89 \protect\cite{wa89}; 
theoretical curves are the same as in Fig.2.} 
\end{center}
\end{figure}

\section{Conclusions.}

There are several conclusions derived from the calculations discussed 
in the article: 

1)~Data of the WA89 experiment on charm production asymmetry can 
be described within the framework of Quark-Gluon String Model 
with the same  asymmetry parameter $a_1=10.$ as  E791 data 
for $\pi^-A$ reaction.

2)~$D^-/D^+$ and $D_s^-/D^+_s$ asymmetries measured with  $\Sigma^-$ 
beam are more sensitive to the weight of charmed quark pairs in the 
quark sea of interacting hyperon ($\delta_{(c,\bar{c})}$=0.01) than 
it could be seen at $\pi^-$ beam interaction 
($\delta_{(c,\bar{c})}$=0.05).

3)~$D_s^-/D_s^+$ asymmetry is higher than $D^-/D^+$ asymmetry because 
strange quark pairs suppressing the asymmetry at $D_s$ production have 
lower weight in quark sea of hyperon than ordinary $d\bar{d}$ pairs 
which cause the suppression of $D^-/D^+$ meson asymmetry. 
 
4)~The both of charmed meson asymmetries are nonzero 
values at $x_F$=0 in these calculations at WA89 energy and diminish 
with the increasing of energy.    

\section{Acknowledgments}

Author would like to express her grattitude to Prof.A.B.Kaidalov, 
Prof.H.-W. Siebert, Prof.J.Pochodzalla and Dr.E.A.Chudakov for the
numerous discussions. She is thankfull to her colleagues of the Particle 
Physics Laboratory of Max-Plank
Institute in Heidelberg for their hospitality during her stay there when 
the basic points of the article were being discussed.
This work was supported by DFG grant $N^{\circ}$  436 RUS 113/332/O(R).

\end{document}